\documentclass[conference]{IEEEtran} 

\makeatletter
\let\NAT@parse\undefined
\makeatother

\usepackage{amsmath,amssymb,amsfonts}
\usepackage[final]{graphicx}
\usepackage{flushend}
\usepackage[numbers,sort&compress]{natbib}

\usepackage{ucs}
\usepackage[utf8x]{inputenc}
\usepackage[usenames,dvipsnames]{xcolor}
\usepackage{tikz}
\usepackage{tkz-tab}
\usepackage{multirow}
\usepackage{latexsym}
\usepackage{amssymb}
\usepackage{amsmath}

\makeatletter
\def\blfootnote{\xdef\@thefnmark{}\@footnotetext}
\makeatother

\begin{document}
\title{\Huge Stochastic Fading Channel Models with Multiple Dominant Specular Components for 5G and Beyond}
\author{
\IEEEauthorblockN{Juan~M.~Romero-Jerez$^{(1)}$, \,F.~Javier~Lopez-Martinez$^{(2)}$,\,Juan P. Pe\~{n}a-Mart\'{i}n$^{(1)}$ and Ali~Abdi$^{(3)}$}
\IEEEauthorblockA{\rm romero@dte.uma.es, fjlopezm@ic.uma.es, jppena@uma.es, ali.abdi@njit.edu}
\IEEEauthorblockA{$^{(1)}$Dpto. de Tecnolog\'ia Electr\'onica, Universidad de M\'alaga, 29071 M\'alaga (Spain).}
\IEEEauthorblockA{$^{(2)}$Dpto. de Ingenier\'ia de Comunicaciones, Universidad de M\'alaga, 29071 M\'alaga (Spain).}
\IEEEauthorblockA{$^{(3)}$Dept. Electrical and Computer Engineering, New Jersey Institute of Technology, Newark, NJ 07102, USA.}
}


\vspace{-4mm}

\maketitle
\begin{abstract}
\blfootnote{\noindent The work of J.M. Romero-Jerez. F. Javier Lopez-Martinez and J.P. Pe\~na-Martin was funded by the Spanish Government
and the European Fund for Regional Development FEDER (project TEC2017-87913-R).
\\
\indent The work of Ali Abdi was supported in part by the National Science Foundation (NSF), grant CCF-0830190.\\
\indent This work has been submitted to the IEEE for publication. Copyright may be transferred without notice, after which this version may no longer be accessible.
}
We introduce a comprehensive statistical characterization of the multipath wireless channel built as a superposition of a number of scattered waves with random phases. We consider an arbitrary number $N$ of specular (dominant) components plus other diffusely propagating waves. Our approach covers the cases on which the specular components have constant amplitudes, as well as when these components experience random fluctuations. These propagation scenarios are found in different use cases of 5G networks, as well as in the context of large intelligent surface based communications. We show that this class of fading models can be expressed in terms of a continuous mixture of an underlying Rician (or Rician shadowed) fading model, averaged over the phase distributions of the specular waves. It is shown that the fluctuations of the specular components have a detrimental impact on performance, and the best performance is obtained when there is only one specular component.
\end{abstract}


\IEEEpeerreviewmaketitle

\section{Introduction}
\label{intro}
With the advent of every new generation of mobile communications, the need for providing high data rates with a low latency and a high reliability pushes the very limits of communication systems and techniques as we know them. The huge variety of use cases considered for 5G \cite{Osseiran2014}, which include device-to-device, machine-to machine and vehicular communications, among many others, entails that such performances are attained in rather dissimilar situations. Because the nature of the wireless channel strongly depends on the operation environment, an accurate characterization of these new wireless propagation scenarios is required.

There has been a great deal of effort on the development of advanced channel models in the context of 5G, aiming to incorporate the geometry of transmitter, receiver and the environment into the channel model \cite{Haneda2016,Hur2016}. This approach, which combines geometric aspects together with stochastic channel modeling, is useful to recreate realistic propagation conditions in a synthetic form. However, due to the large amount of parameters involved in their definition, their use for predicting the behavior of wireless communication systems operating in these environments is far from practical.

In the most general set-up, the received radio signal is formed by the superposition of a set of individual waves, each of which may have a different amplitude and phase \cite{Beckmann1962}. With this formulation, and based on the assumption of a sufficiently large number of received waves, the central limit theorem (CLT) applies and the classical Rayleigh and Rician fading models emerge. In case that such condition does not hold, the distribution of the received signal largely differs from a Gaussian-like distribution. The statistical characterization of the wireless channel where the CLT does not fully apply becomes relevant again today, as in the context of mmWave the diffuse scattering is reduced and only a finite number of multipath components arrives at the receiver \cite{Niu2015}. Similarly, the use of large-intelligent surfaces to modify at will the amplitudes and phases \cite{Subrt2012,Wu2018} of the scattered waves to improve the system performance also justifies the need for a deeper knowledge of the statistics for the equivalent channel.

The statistical characterization of the distribution of the received radio signal is related to one of the key problems in communication theory: the distribution of the sum of $N$ random phase vectors, which is also equivalent to the random walk problem in general statistics. While this problem has been addressed \cite{Slack1946,Rice1974,Simon1985,Helstrom1999} by some of the most reputed communication theorists, its inherent complexity makes it very challenging to characterize its chief statistics. In the context of stochastic channel modeling, it is worth mentioning the pioneering works in \cite{Durgin2000,Abdi2000}, which set the foundations for later advances in the field \cite{Durgin2002,Rao2015,Romero2017}. From a practical perspective, manageable (although in integral form) expressions for the probability density function (PDF) and the cumulative distribution function (CDF) are only available for $N=2$, giving raise to the popular two wave with diffuse power (TWDP) fading model \cite{Durgin2002}. However, as a larger number of specular components is considered, i.e., $N>2$, the evaluation of the PDF and CDF becomes rather hard to be numerically evaluated \cite{Rice1974}. Expansions based on Laguerre polynomial series \cite{Chai2009}, multivariate hypergeometric functions or using the Hankel transform (which requires an improper integration of a highly oscillatory function) are no exception \cite{Chun2018}.

In this paper, we present a new approach to the characterization of wireless channels built as a superposition of $N$ specular waves plus a diffuse component. By conveniently expressing the received signal power in terms of an underlying conditional Rician-like distribution, its PDF, CDF and moment generating function (MGF) are obtained for arbitrary $N$. The resulting expressions require for the evaluation of definite integrals (between $0$ and $2\pi$) of a smooth integrand. This approach has additional benefits from a performance analysis viewpoint, and allows for easily incorporating a random fluctuation on the amplitude of the specular waves, thus generalizing the class of fading models defined in \cite{Romero2017} for arbitrary $N$.

\section{System Model}
\label{analysis NWDP}


The signal in a wireless multipath fading channel can be modeled by the superposition of a set of $N$ dominant waves, referred to as specular components, to which other $M$ diffusely propagating waves are added \cite{Durgin2002}. The received signal can be expressed as 
\begin{equation}
\label{eq:01}
R \exp (j\phi )=\sum\limits_{i = 1}^N  a_i \exp \left( {j\theta _i } \right) + \underbrace{\sum_{i=1}^M A_i  \exp \left( {j\phi _i } \right)}_{A_d=X + jY},
\end{equation}
where $ a_i \exp \left( {j\theta _i } \right)$ represents the \emph{i-th} specular component, which is assumed to have a constant amplitude $a_i$ and a uniformly distributed random phase $\theta _i $, such that $\theta _i \sim \mathcal{U}[0,2\pi)$, where the random phase variables of each specular component are assumed to be statistically independent. Under the assumption that the diffuse received signal component is due to the combined reception of numerous weak, independently-phased scattered waves, then the CLT applies \emph{for this component} and hence we can approximate the last term in \eqref{eq:01}, i.e., $A_d$, as a complex Gaussian random variable, such that $X,Y \sim \mathcal{N}(0,\sigma^2)$. Let $\Omega_0$ denote the average power of the diffuse component. Thus, we can write $E\{|A_d|^2\}=\Omega_0=2\sigma^2$, where $E \left[ \cdot \right]$ denotes the expectation operator.


We note that the model described in (\ref{eq:01}) includes the Rayleigh fading model as a special case for $N=0$, i.e., no specular component is present. For  $N=1$, i.e., a single dominant specular component, we have the Rician fading model. The case in which there are two dominant specular components ($N=2$) is usually referred to as the TWDP fading model, originally proposed by Durgin, Rappaport and de Wolf \cite{Durgin2002}. Our aim will be pursuing the statistical characterization of the model in \eqref{eq:01} for arbitrary $N$. For consistence with the usual nomenclature in the context of fading channel modeling, this will be referred to as the $N$-Wave with Diffuse Power (NWDP) fading model.

\section{Statistical characterization of NWDP fading}
\subsection{Calculation of the PDF and CDF}
\label{PDF NWSP}
Let us define the superposition of the specular components as
\begin{equation}
\label{eq:13}
B_N \exp (j\Psi _N ) \triangleq \sum\limits_{i = 1}^N {a_i } \exp (j\theta _i ),
\end{equation}
so that \eqref{eq:01} becomes
\begin{equation}
\label{eq:01b}
R \exp (j\phi )=B_N \exp (j\Psi _N )+ A_d.
\end{equation}
Note that according to \cite{Scire1968} the distribution of $R$ is independent of the distribution of $\Psi _N$, thanks to the circular symmetry of $A_d$. Conditioned on $B_N$, we have that $R$ follows a Rician distribution, and its PDF will be given by
\begin{equation}
\label{eq:14}
f_{R|B_N } (r|B_N ) = \frac{{2r}}
{{\Omega _0 }}\exp \left( { - \frac{{r^2  + B_N^2 }}
{{\Omega _0 }}} \right)I_0 \left( {\frac{{2B_N r}}
{{\Omega _0 }}} \right),
\end{equation}
where $I_0(\cdot)$ is the zeroth-order modified Bessel function of the first kind.

Let us define
\begin{equation}
\label{eq:001}
\begin{split}
P_N  \triangleq B_N^2  = \left( {\sum\limits_{i = 1}^N {a_i } \cos (\theta _i )} \right)^2  + \left( {\sum\limits_{i = 1}^N {a_i } \sin (\theta _i )} \right)^2,
\end{split}
\end{equation}
which can be rewritten, with the help of the multinomial theorem, as
\begin{equation}
\label{eq:002}
P_N  = \Omega _N + 2\sum\limits_{\Delta _N} {a_i a_k } \cos (\theta _i  - \theta _k )
\end{equation}
with $
\Delta _N  = \left\{ {(i,k):i < k,\;i = 1...N - 1,\;k = 2...N} \right\}
$, and where $\Omega _N$ denotes the total average power of the specular components, verifying
\begin{equation}
\label{eq:20}
\Omega _N  \triangleq E \left[ {P_N } \right] = \sum\limits_{i = 1}^N {a_i^2 }.
\end{equation}

Let us define the power envelope of the received signal $U \triangleq R^2$. The PDF of $U$ will be obtained by averaging over $P_N$ as
\begin{equation}
\label{eq:15}
f_U (u) = E_{P_N } \left[ {\frac{1}
{{\Omega _0 }}\exp \left( { - \frac{{u + P_N }}
{{\Omega _0 }}} \right)I_0 \left( {\frac{2}
{{\Omega _0 }}\sqrt {P_N u} } \right)} \right],
\end{equation}
and the CDF will be given by
\begin{equation}
\label{eq:15b}
F_U (u) = 1 - E_{P_N } \left[ {Q\left( {\sqrt {\frac{{2P_N }}
{{\Omega _0 }}} ,\sqrt {\frac{{2u}}
{{\Omega _0 }}} } \right)} \right],
\end{equation}
where $Q(\cdot,\cdot)$ is the first-order Marcum $Q$-function. Similarly, and leveraging the asymptotic approximation for the Rician CDF in \cite{Wang03}, we can also obtain an asymptotic expression for \eqref{eq:15b}, as the total average power $\Omega _N +\Omega _0 \rightarrow\infty$
\begin{equation}
\label{eq:15c}
F_U (u) \approx \frac{u}{\Omega_0} \cdot E_{P_N } \left[ \exp\left(-\frac{P_N}{\Omega_0}\right)\right].
\end{equation}
From \eqref{eq:15c} we see that the diversity order (i.e. the exponent of $u$) is one $\forall N$, and the power offset depends on the power of the dominant component in a very simple form. These are new results in the literature to the best of our knowledge.

Without any loss of generality, we will assume that $\theta _1=0$. From (\ref{eq:001}) and (\ref{eq:002}), averaging over $P_N$ is equivalent to averaging over $\theta _2,...,\theta _N$, i.e., $N-1$ nested integrals in the interval $[0,2\pi)$ need to be computed. This kind of computation is rather common in communication theory \cite{AlouiniBook} and actually does not pose any numerical challenge because the integrand is a continuous bounded function and the integration interval is finite. In addition, the case $N>4$ will rarely need to be considered, as the proposed model will rapidly converge to either Rayleigh or Rician fading by virtue of the central limit theorem \cite{Durgin2002,Chun2018}. As an additional benefit of this approach, we will now show that the MGF expression can be manipulated to eliminate one nested integral, so that $N-2$ integrations need to be performed, i.e. a finite-range single integral will need to be computed for $N=3$ and an easy-to-compute double integral will be required for $N=4$.

\subsection{Calculation of the MGF}
\label{MGF NWSP}

The MGF of the envelope power $U$, conditioned on the instantaneous power of the specular components, $P_N$, can be calculated as
\begin{equation}
\label{eq:16}
\begin{split}
  & M_N (s|P_N ) = \cr
  & {\frac{1}
{{\Omega _0 }}\int_0^\infty  {\exp \left( {su - \frac{{u + P_N }}
{{\Omega _0 }}} \right)I_0 \left( {\frac{2}
{{\Omega _0 }}\sqrt {P_N u} } \right)du} }   \cr
  & \quad \quad  =   {\frac{1}
{{1 - \Omega _0 s}}\exp \left( {\frac{{P_N s}}
{{1 - \Omega _0 s}}} \right)} .
\end{split}
\end{equation}

It has been shown in \cite{Chen2011} that $P_N$ can be computed recursively as
\begin{equation}
\label{eq:24}
P_N  = P_{N-1}  + a_N^2  + 2a_N \sqrt {P_{N-1} } \cos \left( {\theta _N  - \Psi _{N - 1} } \right),
\end{equation}
where $P_0=0$.
Therefore, we can express the MGF of $U$ conditioned on $P_{N-1}$, $\theta _N$ and $\Psi _{N - 1}$ as
\begin{equation}
\label{eq:25}
\begin{split}
  & M_N (s|P_{N-1} ,\theta _N ,\Psi _{N - 1} ) = \frac{1}
{{1 - \Omega _0 s}}  \cr
  &  \cdot \exp \left( {\frac{{\left( {P_{N-1}  + a_N^2  + 2a_N \sqrt {P_{N-1} } \cos \left( {\theta _N  - \Psi _{N - 1} } \right)} \right)s}}
{{1 - \Omega _0 s}}} \right).
\end{split}
\end{equation}
Considering that
$
I_0 \left( x \right) = 1/(2\pi )\int_0^{2\pi } {\exp \left( {x\cos \left( {\beta  - \xi } \right)} \right)} d\xi
$,
the dependency on $\theta _N$ and $\Psi _{N - 1}$ can be eliminated in (\ref{eq:25}) by computing
\begin{equation}
\label{eq:26}
M_N (s|P_{N-1} ) = \frac{1}
{{2\pi }}\int_0^{2\pi } {M_N (s|P_{N-1} ,\theta _N ,\Psi _{N - 1} )} d\theta _N ,
\end{equation}
yielding
\begin{equation}
\label{eq:27}
\begin{split}
  & M_N (s|P_{N-1} ) = \frac{1}
{{1 - \Omega _0 s}}\exp \left( {\frac{{\left( {P_{N-1}  + a_N^2 } \right)s}}
{{1 - \Omega _0 s}}} \right)  \cr
  & \quad \quad \quad \quad \quad  \times I_0 \left( {\frac{{2a_N \sqrt {P_{N-1} } s}}
{{1 - \Omega _0 s}}} \right).
\end{split}
\end{equation}
The unconditional MGF will thus be given by:
\begin{equation}
\label{eq:28}
\begin{split}
  & M_N (s) = \frac{1}
{{1 - \Omega _0 s}}  \cr
  &  \times E_{P_{N-1} } \left[ {\exp \left( {\frac{{\left( {P_{N-1}  + a_N^2 } \right)s}}
{{1 - \Omega _0 s}}} \right)I_0 \left( {\frac{{2a_N \sqrt {P_{N-1} } s}}
{{1 - \Omega _0 s}}} \right)} \right].
\end{split}
\end{equation}
The recursive expression of $P_N$ given in (\ref{eq:24}) was helpful to obtain (\ref{eq:28}), however, the practical use of (\ref{eq:24}) in order to perform the expectation operation in (\ref{eq:28}) can be cumbersome. A more practical expression of $P_N$ was given in (\ref{eq:002}). We note that the MGF of the pure $N$-Ray model (i.e. withouth diffuse component) is obtained as a by-product by setting $\Omega_0=0$.

For $N = 1$, (\ref{eq:25}) collapses to the well know MGF of the squared Rician distribution:
\begin{equation}
\label{eq:29}
M_1 (s) = \frac{1}
{{1 - \Omega _0 s}}\exp \left( {\frac{{a _1^2 s}}
{{1 - \Omega _0 s}}} \right).
\end{equation}
When $N = 2$, since $P_1  = a_1^2$ is a constant, (\ref{eq:28}) results in
\begin{equation}
\label{eq:30}
\begin{split}
M_2 (s) = \frac{1}
{{1 - \Omega _0 s}}\exp \left( {\frac{{\left( {a_1^2  + a_2^2 } \right)s}}
{{1 - \Omega _0 s}}} \right)I_0 \left( {\frac{{2a_1 a_2 s}}
{{1 - \Omega _0 s}}} \right),
\end{split}
\end{equation}
which is the MGF of a squared TWDP distribution, recently derived in \cite{Rao2015}.
When $N = 3$, the MGF can be efficiently calculated by performing a finite-range integral in $[0,2\pi)$, as in this case we can write
\begin{equation}
\label{eq:31}
\begin{split}
  & M_3 (s) = \frac{1}
{{1 - \Omega _0 s}}  \cr
  & \quad \quad  \times E_{\theta _2 } \left[ {\exp \left( {\frac{{\left( {P_2  + a_3^2 } \right)s}}
{{1 - \Omega _0 s}}} \right)I_0 \left( {\frac{{2a_3 \sqrt {P_2 } s}}
{{1 - \Omega _0 s}}} \right)} \right],
\end{split}
\end{equation}
where $P_2  = a_1^2  + a_2^2  + 2a_1 a_2 \cos \theta _2 $.

Similarly, for $N=4$, the MGF can be computed by performing a double finite-range integral in $[0,2\pi)$, as in this case we have
\begin{equation}
\label{eq:32}
\begin{split}
  & M_4 (s) = \frac{1}
{{1 - \Omega _0 s}}  \cr
  & \quad \quad  \times E_{\theta _2 ,\theta _3 } \left[ {\exp \left( {\frac{{\left( {P_3  + a_4^2 } \right)s}}
{{1 - \Omega _0 s}}} \right)I_0 \left( {\frac{{2a_4 \sqrt {P_3 } s}}
{{1 - \Omega _0 s}}} \right)} \right],
\end{split}
\end{equation}
where $
P_3  = a_1^2  + a_2^2  + a_3^2  + 2a_1 a_2 \cos \theta _2  + 2a_1 a_3 \cos \theta _3  + 2a_2 a_3 \cos (\theta _2  - \theta _3 )
$.

\section{A generalization of NWDP fading}
\label{analysis FNR}

The specular components in the general model in (\ref{eq:01}) have constant amplitudes. We must here note that variations in the amplitude of the dominant specular components have been considered in some scenarios and validated with field measurements. These are the cases of the Rician shadowed fading model \cite{Abdi2003} or the Fluctuating Two-Ray (FTR) fading model \cite{Romero2017}, where
the amplitudes of the specular components are assumed to be modulated by a normalized gamma random variable. Generalizing this approach, we can write:
\begin{equation}
\label{eq:005}
R \exp (j\phi )=\sum\limits_{i = 1}^N  \sqrt{\zeta} a_i \exp \left( {j\theta _i } \right) + X + jY,
\end{equation}
where $\zeta$ is a unit-mean Gamma distributed random variable with PDF
\begin{equation}
\label{eq:006}
f_\zeta  \left( u \right) = \frac{{m^m u^{m - 1} }}
{{\Gamma \left( m \right)}}e^{ - mu} .
\end{equation}
Note that we are considering the same fluctuation for the specular components, which is a natural situation in different wireless scenarios on which the scatterers are in the vicinity of the transmitter and/or the receiver, as discussed in \cite{Romero2017}.

The wireless channel model given in (\ref{eq:005})-(\ref{eq:006}) for the particular case when ${N=1}$ corresponds to the Rician shadowed fading model \cite{Abdi2003} and when ${N=2}$ corresponds to the FTR fading model \cite{Romero2017}. The general proposed model will be referred to as the Fluctuating $N$-Ray (FNR) fading model and, to the authors' knowledge, its statistical characterization has not been presented for $N>2$ in a tractable form \cite{Chun2018}.

\subsection{Calculation of the PDF and CDF}
\label{PDF FNR}
Let us define $P_N$ as in (\ref{eq:001}) and (\ref{eq:002}).
Conditioning on $P_N$, and following the same rationale as in the previous Section, the FNR fading model reduces to a conditional Rician shadowed fading model. Thus, using \cite[eq. (6)]{Abdi2003} the distribution of the received power envelope of the FNR model will be given by

\begin{equation}
\label{eq:007}
\begin{split}
  & f_U (u) = E_{P_N } \left[ {\left( {\frac{{\Omega _0 m}}
{{\Omega _0 m + P_N }}} \right)^m \frac{1}
{{\Omega _0 }}\exp \left( { - \frac{u}
{{\Omega _0 }}} \right)\;} \right.  \cr
  & \quad \quad  \times \left. {\;{}_1 F_1 \left( {m;1;\frac{{P_N u}}
{{\Omega _0 \left( {\Omega _0 m + P_N } \right)}}} \right)} \right], \cr
\end{split}
\end{equation}
where $_1F_1$ is the confluent hypergeometric function of the first kind.
With the help of \cite[eq. (8)]{Paris2010}, the CDF can be written
\begin{equation}
\label{eq:007b}
\begin{split}
  & F_U (u) = \frac{u}
{{\Omega _0 }}E_{P_N } \left[ {\left( {\frac{{\Omega _0 m}}
{{\Omega _0 m + P_N }}} \right)^m } \right.  \cr
  & \quad  \times \left. {\Phi _2 \left( {1 - m;m;2; - \frac{u}
{{\Omega _0 }}; - \frac{{mu}}
{{\Omega _0 m + P_N }}} \right)} \right], \cr
\end{split}
\end{equation}
where $\Phi _2$ is the bivariate confluent hypergeometric function defined in \cite[p. 34, (8)]{Srivastava1985}. Note that when the fading parameter $m$ is a positive integer, then the PDF and CDF of the Rician shadowed model are given in terms of a finite sum of powers and exponentials \cite{Paris2010}. Hence, the consideration of a random fluctuation in the specular components brings additional benefits from a practical perspective, as the numerical integration of these functions is always simpler than its deterministic counterpart.

An asymptotic expression for the CDF similar to that in \eqref{eq:15c} can be obtained using \cite{Paris2014} as
\begin{equation}
\label{eq:007c}
F_U (u) \approx \frac{u}{\Omega_0} \cdot E_{P_N } \left[ \left(\frac{m}{\frac{P_N}{\Omega_0}+m} \right)^m \right].
\end{equation}
Again, the diversity order of the FNR fading model is one for arbitrary $N$.

\subsection{Calculation of the MGF}
\label{MGF FNR}

Conditioning on $P_N$, the MGF can be written with the help of \cite[eq. (7)]{Abdi2003}\footnote{Which can be written in a more compact form with respect to the expression given in the reference.} as
\begin{equation}
\label{eq:008}
M_N \left( {s|P_N } \right) = \frac{{m^m \left( {1 - \Omega _0 s} \right)^{m - 1} }}
{{\left( {m - \left( {m\Omega _0  + P_N } \right)s} \right)^m }}.
\end{equation}
Let us define
\begin{equation}
\label{eq:009}
\beta \left( {P_{N-1} } \right) \triangleq m\Omega _0  + P_{N-1}  + a_N^2 .
\end{equation}
Introducing (\ref{eq:24}) into (\ref{eq:008}), we can write
\begin{equation}
\label{eq:010}
\begin{split}
  & M_N (s|P_{N-1} ,\theta _N ,\Psi _{N - 1} ) =   \cr
  & \frac{{m^m \left( {1 - \Omega _0 s} \right)^{m - 1} }}
{{\left( {m - \left( {\beta \left( {P_{N-1} } \right) + 2a_N \sqrt {P_{N-1} } \cos \left( {\theta _N  - \Psi _{N - 1} } \right)} \right)s} \right)^m }} .\cr
\end{split}
\end{equation}
Let us now assume that parameter $m$ takes integer values for the sake of simplicity. Averaging with respect to the random variable $\theta_N$ and noticing that the function given in (\ref{eq:010}) is periodic with period $2\pi$ with respect to this variable, with the help of \cite[eq. 3.661.4]{Gradstein2007}, we can obtain
\begin{equation}
\label{eq:011}
\begin{split}
  & M_N (s) = E_{P_{N-1} } \left[ {\frac{{m^m \left( {1 - \Omega _0 s} \right)^{m - 1} }}
{{\left( {\sqrt {\left[ {m - \beta \left( {P_{N-1} } \right)s} \right]^2  - 4a_N^2 P_{N-1} s^2 } } \right)^{m} }}} \right.  \cr
  & \quad \quad \quad  \times \left. {P_{m-1} \left( {\frac{{m - \beta \left( {P_{N-1} } \right)s}}
{{\sqrt {\left[ {m - \beta \left( {P_{N-1} } \right)s} \right]^2  - 4a_N^2 P_{N-1} s^2 } }}} \right)} \right], \cr
\end{split}
\end{equation}
where $P_n(\cdot)$ is the Legendre polynomial of degree\footnote{Relaxing the assumption of $m\in\mathbb{Z}^+$ implies that \eqref{eq:011} is expressed in terms of the Legendre function $\mathcal{P}_{m-1}(z)$ \cite{Romero2017}, which is related to the Gauss Hypergeometric function as $\mathcal{P}_{\mu}(z) ={}_2F_1(-\mu,\mu+1;1;\frac{1-z}{2})$.} $n$, which can be written as \cite[p. 775 (22.3.8)]{Abramowitz72}
\begin{equation}
\label{eq:012}
P_n \left( z \right) = \frac{1}
{{2^n }}\sum\limits_{q = 0}^{\left\lfloor {n/2} \right\rfloor } {\left( { - 1} \right)^q } C_q^n z^{n - 2q},
\end{equation}
where $\left\lfloor  \cdot  \right\rfloor $ is the floor function and $C_q^n$ is a coefficient given by
\begin{equation}
\label{eq:012b}
C_q^n  = \left( {\begin{array}{c}
   n  \\
   q  \\
 \end{array} } \right)\left( {\begin{array}{c}
   {2n - 2q}  \\
   n  \\
 \end{array} } \right) = \frac{{\left( {2n - 2q} \right)!}}
{{q!\left( {n - q} \right)!\left( {n - 2q} \right)!}}.
\end{equation}

When $N=1$, we can write
\begin{equation}
\label{eq:013}
M_1 (s) = \frac{{m^m \left( {1 - \Omega _0 s} \right)^{m - 1} }}
{{\left( {m - \left( {m\Omega _0  + a_1^2 } \right)s} \right)^m }},
\end{equation}
while for $N=2$ we have
\begin{equation}
\label{eq:014}
\begin{split}
  & M_2 (s) = \frac{{m^m \left( {1 - \Omega _0 s} \right)^{m - 1} }}
{{\left( {\sqrt {\left[ {m - \left( {m\Omega _0  + a_1^2  + a_2^2 } \right)s} \right]^2  - 4a_1^2 a_2^2 s^2 } } \right)^m }}  \cr
  & \quad \quad  \times P_{m - 1} \left( {\frac{{m - \left( {m\Omega _0  + a_2^2  + a_2^2 } \right)s}}
{{\sqrt {\left[ {m - \left( {m\Omega _0  + a_1^2  + a_2^2 } \right)s} \right]^2  - 4a_1^2 a_2^2 s^2 } }}} \right). \cr
\end{split}
\end{equation}

\section{Numerical results}
\label{numerical}
In this section, we present some illustrative results of the statistics and performance of the channel models previously characterized, aiming to identify the impact of the different parameters. In all cases, Monte Carlo (MC) simulations have been included in the figures through markers in order to confirm the validity of the derived expressions. We define a power ratio parameter similar to the conventional Rician $K$ parameter as $K_{N}\triangleq \frac{{\Omega_N}}{\Omega_0}$, which will be useful in the sequel. 

\subsection{Probability density function}

\begin{figure}[t]
\centering
\includegraphics[width=\columnwidth]{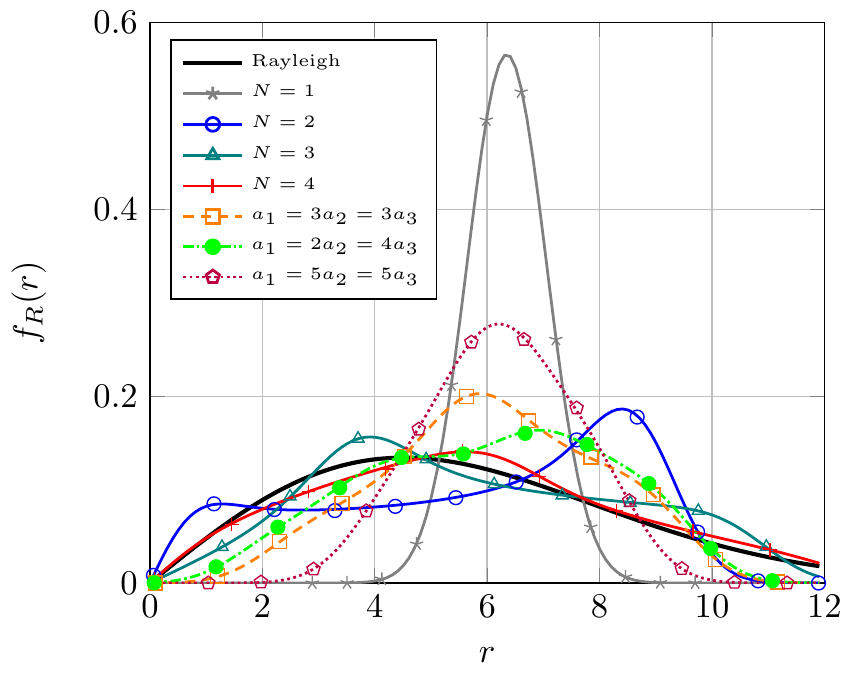}
\caption{Probability density function of the received signal amplitude under NWDP fading, for different numbers of dominant specular waves $N$ and different amplitude configurations. Parameter values are $K_{\rm dB}=16$ dB and $\Omega_0=1$. Solid lines correspond to the balanced amplitude cases. Markers denote MC simulations.}
\label{Figura1}
\end{figure}

\begin{figure}[t]
\centering
\includegraphics[width=\columnwidth]{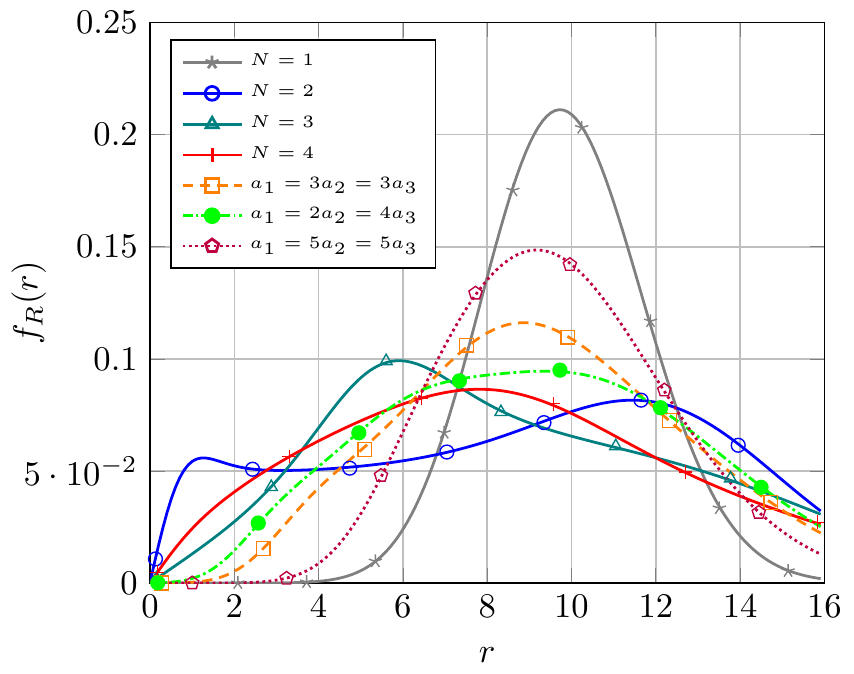}
\caption{Probability density function of the received signal amplitude under FNR fading, for different numbers of dominant specular waves $N$ and different amplitude configurations. Parameter values are $K_{\rm dB}=20$ dB, $\Omega_0=1$ and $m=8$. Solid lines correspond to the balanced amplitude cases. Markers denote MC simulations.}
\label{Figura1b}
\end{figure}

We first represent the PDF of the received signal envelope $R$ in Figs. \ref{Figura1} and \ref{Figura1b}, which are directly obtained from those of $U$ using a simple transformation of random variables, i.e. $f_R(r)=2r\cdot f_U(r^2)$. Fig. \ref{Figura1} corresponds to the NWDP case, on which different parameter configurations have been chosen. We set $K_{\rm dB}=10\log_{10}(K_N)=16$ dB\footnote{A reasonably large value of $K_{\rm dB}$ is required in this setting, as smaller values of $K_{\rm dB}$ would lead the lower-amplitude dominant components to be indistinguishable from the diffuse one.} with $\Omega_0=1$, so that ${E\{|R|^2\}=K_N+1}$, and two different situations: balanced amplitudes of the dominant waves ($a_i=a_j \forall i,j=1\ldots N$), as well as the unbalanced amplitude situation. The balanced case is represented in the figure using solid lines, and the unbalanced case (with $N=3$) is represented using discontinuous lines.

Let us first analyze the balanced scenario: we see that the cases $N=2$ and $N=3$ correspond to very different PDF shapes compared to the Rician case ($N=1$), and may lead to a non-unimodal behavior. For $N=4$ we see that the shape of the PDF is rather different from the Rician case, although it starts to resemble that of a Rayleigh distribution. With regard to the unbalanced case, we see that modifying the relative amplitudes of the dominant waves has a direct impact on the shape of the PDF, and notably on the mode of the distribution. As the unbalance is more noticeable the shape of the PDF tends to a Rician distribution (with a larger variance), because the power of the lower-amplitude components becomes similar to the diffuse component.

Fig. \ref{Figura1b} illustrates the PDF of the received signal envelope under FNR fading, using a similar set of parameters as those in Fig. \ref{Figura1}. We now set a slightly larger $K_{\rm dB}=20$ dB and $m=8$, which corresponds to a moderate fluctuation of the dominant components. We see that the effect of considering a random fluctuation on the dominant components somehow averages out the shape of the PDF compared to the deterministic case (i.e. NWDP). Increasing the number of waves or modifying their relative amplitudes has a similar effect as in the NWDP case.

\subsection{Outage capacity }

The instantaneous channel capacity per unit bandwidth is given by
\begin{equation}
\label{eq:48}
C = \log _2 (1 + \gamma ),
\end{equation}
where $\gamma=U / N_0$ is the instantaneous signal-to-noise ratio (SNR) and $N_0$ is the background noise. We define the outage capacity as the probability that the instantaneous channel capacity $C$ falls below a predefined threshold  $R_S$ (given in terms of rate per unit bandwidth), i.e.,
\begin{equation}
\label{eq:49}
P_{\rm out}  = P\left( {C < R_S } \right) = P\left( {\log _2 (1 + \gamma ) < R_S } \right).
\end{equation}
Therefore
\begin{equation}
\label{eq:50}
P_{\rm out}  = P\left( {\gamma  < 2^{R_S }  - 1} \right) = F_{\gamma}  \left( {2^{R_S }  - 1} \right).
\end{equation}
Thus, the outage capacity probability can be directly calculated from (\ref{eq:15b}) and (\ref{eq:007b}) for the NWDP and FNR fading models, respectively, specialized for $\gamma=2^{R_S } -1$. Note that the average SNR is given by $\overline\gamma=E\{U\}/N_0$.

\begin{figure}[t]
\centering
\includegraphics[width=\columnwidth]{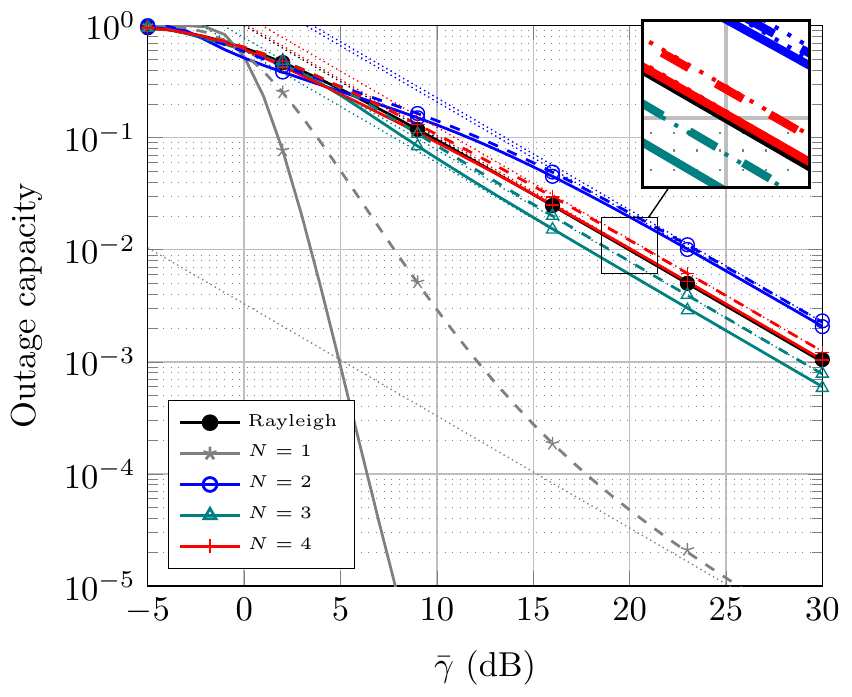}
\caption{Outage capacity vs. average SNR, for different numbers of dominant specular waves $N$. The Rayleigh case is included as a reference (solid black line). Solid colored lines correspond to NWDP fading, and dashed lines correspond to the FNR case. Dotted thin lines are used for the asymptotic approximations using \eqref{eq:15b} and \eqref{eq:007c}. Parameter values are $K_{\rm dB}=14$ dB, $\Omega_0=1$, $m=5$ and $R_S=1$ bps/Hz. Markers denote MC simulations}
\label{Figura2}
\end{figure}

The outage capacity is evaluated in Fig. \ref{Figura2}, aiming to understand the impact of the number of dominant components in the system performance. The balanced case is considered, and both the exact and the asymptotic outage capacity are represented in the figure. A threshold rate $R_S=1$ bps/Hz is set, and the fading parameter values are $K_{\rm dB}=14$ dB, $\Omega_0=1$ and $m=5$. We see that the best performance is always attained for $N=1$, i.e. the Rician or Rician shadowed cases. As indicated in \cite{Chai2009}, the worst performance is achieved for the case $N=2$; notably, the performance is even worse than in the Rayleigh case. Increasing the number of waves to $N=3$ is beneficial for system performance compared to $N=2$, and for $N=4$ we see a very similar (although slightly worse) performance than in the Rayleigh case. In all instances, the performance under FNR fading is always worse than its NWDP counterpart.

\section{Conclusions}
\label{conc}
We have presented a comprehensive statistical characterization of the multipath wireless channel, when numerous scattered waves (either with constant or random amplitudes) are received. We showed that the case of arbitrary $N$ can be reduced to that of an underlying Rician (or Rician shadowed) channel, which considerably facilitates its analysis. The newly proposed class of fading models is able to modeling propagation conditions
which largely differ from those captured by classical fading models like Rayleigh or Rician.


\bibliographystyle{IEEEtran}
\bibliography{bibjavi}

\end{document}